\documentclass[conference]{IEEEtran}
\IEEEoverridecommandlockouts
\usepackage{cite}
\usepackage{amsmath,amssymb,amsfonts}
\usepackage{algorithmic}
\usepackage{graphicx}
\usepackage{textcomp}
\usepackage{xcolor}
\usepackage{soul}
\usepackage{comment}

\setstcolor{red}
\newif\ifreview
\reviewfalse

\ifreview
\newcommand{\nk}[1]{\textcolor{blue}{#1}}
\newcommand{\nkst}[1]{\st{#1}}
\else 
\newcommand{\nk}[1]{{#1}}
\newcommand{\nkst}[1]{}
\fi

\def\BibTeX{{\rm B\kern-.05em{\sc i\kern-.025em b}\kern-.08em
    T\kern-.1667em\lower.7ex\hbox{E}\kern-.125emX}}
\begin{document}

\title{Implementation and Validation of Mismatch-Tolerant High-Gain OTA with a Calibrated Negative Conductance Load}

\author{\IEEEauthorblockN{Mayank Aggarwal and Naveen Kadayinti}
\IEEEauthorblockA{Department of Electrical, Electronics and Communication Engineering \\
Indian Institute of Technology Dharwad, Karnataka - 580011, India\\
Email: \{ee23ms003,naveen\}@iitdh.ac.in}
}

\maketitle

\begin{abstract}
This paper presents the circuit-level implementation of a mismatch-tolerant high-gain operational transconductance amplifier (OTA) using automatic negative-conductance calibration in 65-nm CMOS. The technique enhances OTA gain by compensating for its finite output conductance using a cross-coupled transistor pair, with an on-chip calibration loop that eliminates the need for a replica circuit or external tuning. \nk{The calibration routine increases the gain by driving the steady state error of the OTA in a unity gain feedback loop toward zero.} The implementation addresses practical challenges involving comparator offset, DAC design, and subtractor gain and offset. A 6-bit capacitive DAC (CDAC)-based comparator offset calibration scheme enables reliable detection of the sub-millivolt \nk{steady state} error required for high OTA gain. Post-layout Monte Carlo simulations show that the comparator offset standard deviation is reduced from 4.11 mV to 132~$\mu$V, with a maximum observed post-calibration offset of 625~$\mu$V.\nkst{ The negative-conductance control DAC provides a 24-mV calibration range.} The complete calibration circuitry is \nkst{physically }implemented and evaluated through post-layout simulations. 
\end{abstract}


\section{Introduction}

\nkst{\it A. Motivation}

Operational transconductance amplifiers (OTAs) are widely used as fundamental building blocks in analog and mixed-signal circuits such as Gm-C filters, integrators, amplifiers, and data converters. \nk{The analyses and design flow of circuits with OTA's usually assume OTA's as ideal transconductors with infinite output impedance}. In \nkst{practical OTAs}\nk{practice}, however, the output impedance \nk{of OTAs} is finite\nkst{, unlike the ideal case where it is assumed to be infinite}. This finite output impedance introduces loading between circuit nodes, reduces effective gain, and \nk{invariably shifts} \nkst{can shift}the overall circuit's poles and zeros \nk{from the design values}, causing deviations from the desired transfer characteristics. \nkst{In complex circuit designs, where the number of OTAs are more, more error will be observed in the performance characteristics, as shown in [unpublished ISCAS paper] with an example of a 3rd-order Bessel Gm-C filter.}\nk{This problem of low output impedance becomes worse with technology scaling, due to degrading analog performance of transistors~\cite{1,2}}. \nkst{Moreover, output impedance continues to decrease as we move toward advanced node technologies, as explained in ?cite{1} and ?cite{2}.}Therefore, \nk{novel OTA circuits that feature} high-gain \nkst{OTAs}\nkst{along with high output impedance}\nk{in scaled technologies are essential} \nkst{are desirable}to minimize these effects.

\nk{One of the classic ways of achieving high output impedance is by using cascode or folded cascode OTAs. These, however, come with constraints of headroom in low supply voltage designs. A technique which is more widely adopted in recent years is the use of negative impedance (realized using controlled positive feedback) to cancel the output impedance, thereby resulting in high output impedance~\cite{marinaik2023,nauta1992, Gain1, mondal2015, Gain2}.}\nkst{Achieving high gain in practical implementations introduces additional challenges such as limited voltage headroom in cascode topology, reduced output swing, or bandwidth deterioration, as in cascading. Another useful approach involves applying positive feedback, often implemented with an X-CP (cross-coupled pair), to effectively increase the output resistance and hence the gain. To understand this, let's look back again at Figure~1.} 
\begin{figure}[h!]
   \centering
   \includegraphics[scale=0.12]{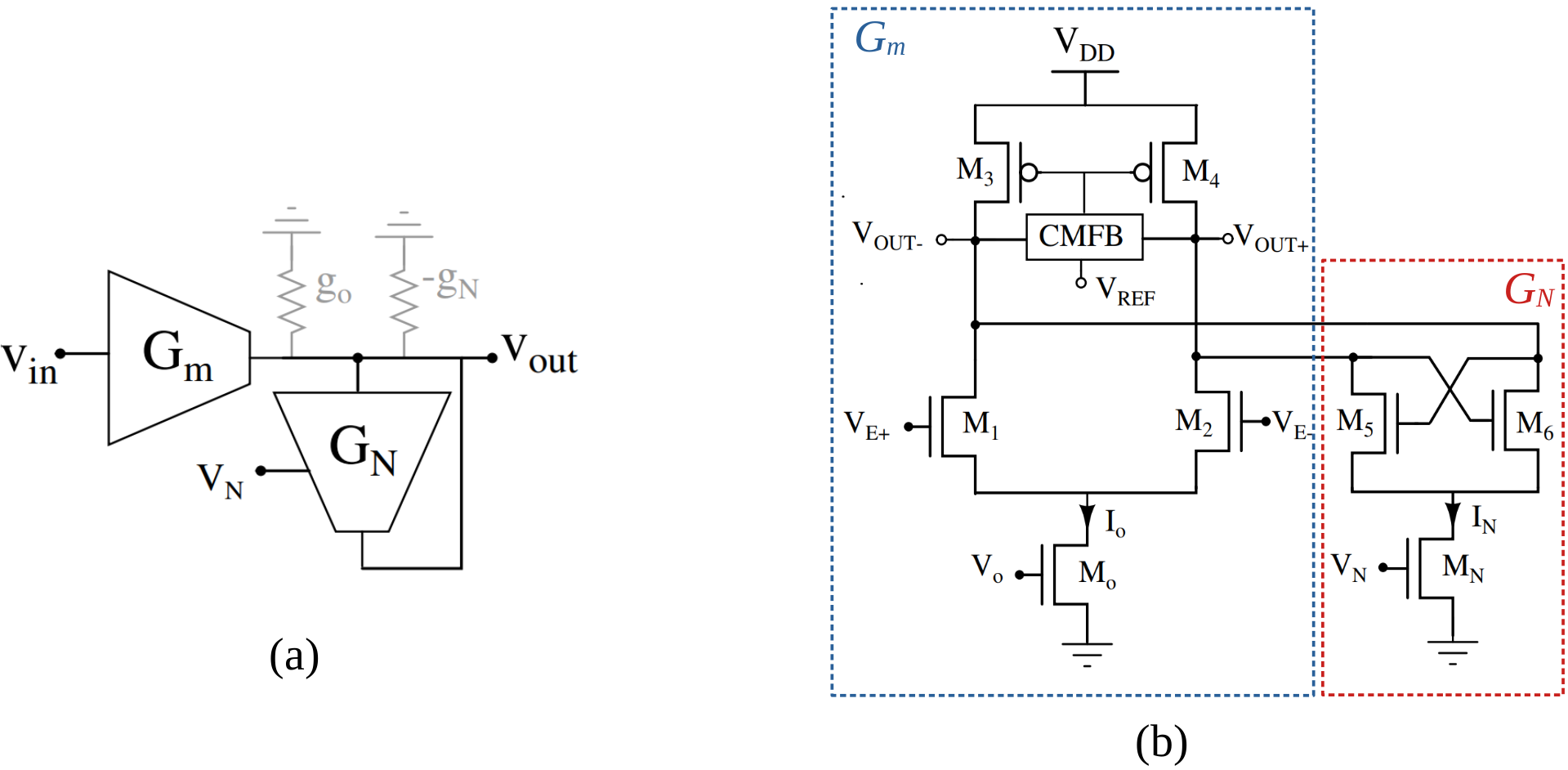}
   \caption{ (a) Output impedance cancellation with the help of negative conductance ($-g_{N}$) connected at load of an OTA with a practical output conductance $g_{o}$ \cite{marinaik2023} (b) Realization of negative conductance from cross-coupled MOS-pair $G_{N}$ \cite{nauta1992} (Red) where OTA has transconductance $G_{m}$ (Blue) with $V_{E}$ as input to the OTA comprised of $M_0 - M_4$ \cite{marinaik2023}.}
   \label{2.2}
   \vspace{-0.2cm}
\end{figure}
\nk{The principle of negative impedance based high gain OTA is illustrated in Fig.~\ref{2.2}.} Since the gain is decreased due to the \nkst{finite}\nk{non-zero} output \nkst{impedance}\nk{conductance} of the OTA, \nkst{g?textsubscript{O}, }the gain can be increased by cancelling g\textsubscript{O} by connecting a negative conductance ($-g_{N}$) in parallel with it, as shown in Fig.~\ref{2.2}(a). Positive\nkst{ly} feedback transconductor ($G_{N}$) is realized using a cross-coupled pair and shown in Fig.~\ref{2.2}(b). \nkst{It can easily be deduced from Figure~?ref{2.2} (a) that}The net output conductance after connecting the negative impedance becomes \cite{marinaik2023}:
\begin{equation}
g_{ON} = g_O - g_N .
\end{equation}
\nkst{And the}\nk{The} DC gain will be \cite{marinaik2023}:
\begin{equation}
A_{ON} = \frac{G_m}{g_{ON}} .
\end{equation}
\nk{Driving $g_{ON}$ to zero can in principle make the gain infinite. However, in practice mismatch and the time constant of the positive feedback loop limit the gain to finite values.}

\nkst{As seen from Figure~?ref{2.2} (b), it should be clear that the value of $g_{N}$ is determined by the value of $V_{N}$, which reduces $g_{ON}$ and hence increases $A_{ON}$.}

Many reported architectures have increased the gain of the OTA by using this method. In \cite{nauta1992}, the authors have achieved a gain of 46~dB using two inverters in a back-to-back configuration. As a comparison to this, \cite{Gain1} used a similar topology and achieved a gain of 80~dB. Both \cite{nauta1992} and \cite{Gain1}\nkst{, uses a power supply of 10V. Such high supply voltages and power consumption are unsuitable for integrated CMOS circuits implemented in advanced technology nodes, including the APF or TTD elements targeted in this work.} \nk{were early demonstrations of this concept, which used separate external supply voltage for the positive feedback circuit that was tuned manually for calibrating its conductance.} A similar approach is used in \cite{Gain2}, achieving a gain of 83~dB with a cross-drain-coupled differential amplifier. Even though \cite{nauta1992}, \cite{Gain1}, and \cite{Gain2} report the cancellation of output conductance, none of them have investigated a way to implement the cancellation across PVT without any external intervention. One of the major challenges in such a technique is to generate and fix the bias voltage, $V_N$, for the positive feedback conductance (or resistance) used for the impedance cancellation (refer to Figure~1).
One of the methods for generating this bias is \nkst{(as discussed in ?cite mondal2015) }to incorporate a replica of the OTA, which will track the output conductance, $g_{ds}$~\cite{mondal2015}. \nkst{But since there is a replica, there will be mismatches, which will directly lead to improper cancellation of $g_{ds}$, compromising the technique's efficiency, wherein the authors have achieved a minimum gain of 42~dB across all corners.}\nk{Here, any mismatch between the OTA in the circuit in question and the replica will result in 
imperfect cancellation. The authors of~\cite{mondal2015} report a minimum gain of 42~dB across all corners.}
A similar technique is reported in \cite{marinaik2023}, which is based on the same principle of impedance cancellation, \nk{but with the calibration is done in the foreground without requiring any replica. Here}\nkst{ where} the error voltage is sensed with the OTA in a unity-gain feedback, and the gain is increased by adjusting the negative impedance to reduce the error voltage to zero. This in-line high-gain calibration technique does not require a replica, making it mismatch-tolerant. This technique reported a minimum gain of 50~dB across PVT in \cite{marinaik2023}. \nk{In this technique, the achievable gain of the OTA is limited by the accuracy of the circuit that senses the steady state error of the OTA in unity gain feedback.}

\subsection{{OTA Calibration Principle}}

The basic principle of the technique for determining $V_N$, as shown in Fig.~2, is to reduce the steady state error (= $v_e$) of the OTA configured in unity gain feedback to zero. Ideally, this increases the DC gain of the OTA to infinity.

\begin{figure}[h!]
   \centering
   \includegraphics[scale=0.12]{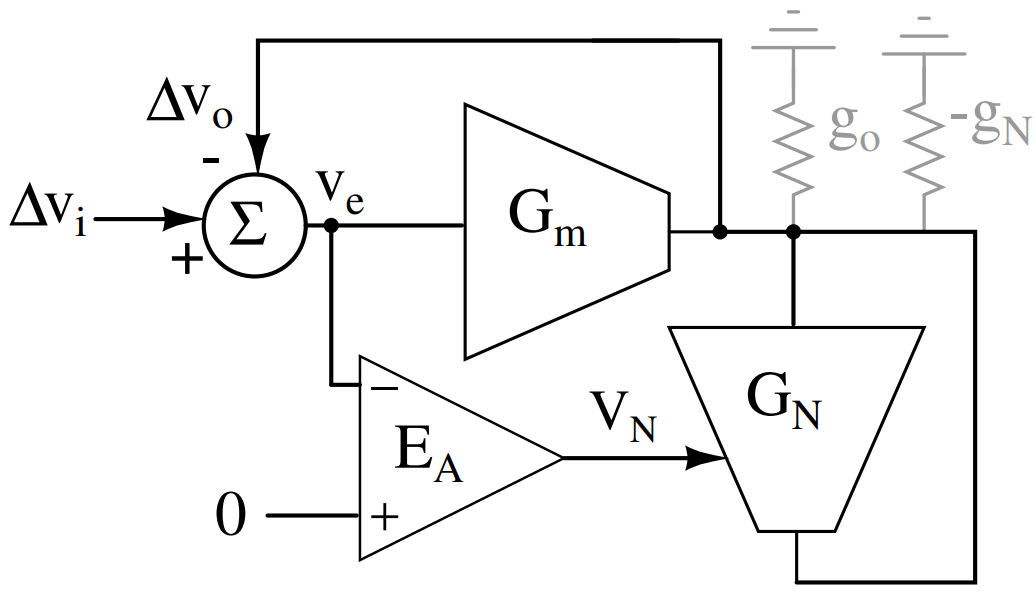}
   \caption{ $V_{N}$ detection by reducing the steady state error voltage of OTA $v_{e}$ \cite{marinaik2023}.}
   \label{2.3}
   \vspace{0.2cm}
\end{figure}

Consider Fig.~\ref{2.3}, where the input is an incremental voltage $\Delta v_i$. \nkst{The output voltage $\Delta v_o$ and error voltage $v_e$ can be expressed as follows~?cite{marinaik2023}:}
%
%
%
%
The error voltage can then be written as:

\begin{equation}
v_e = \Delta v_i \frac{g_{ON}}{G_m + g_{ON}}
\label{eq:2.3}\ ,
\end{equation}

\begin{figure}[h!]
   \centering
   \vspace{0.2cm}
   \includegraphics[scale=0.12]{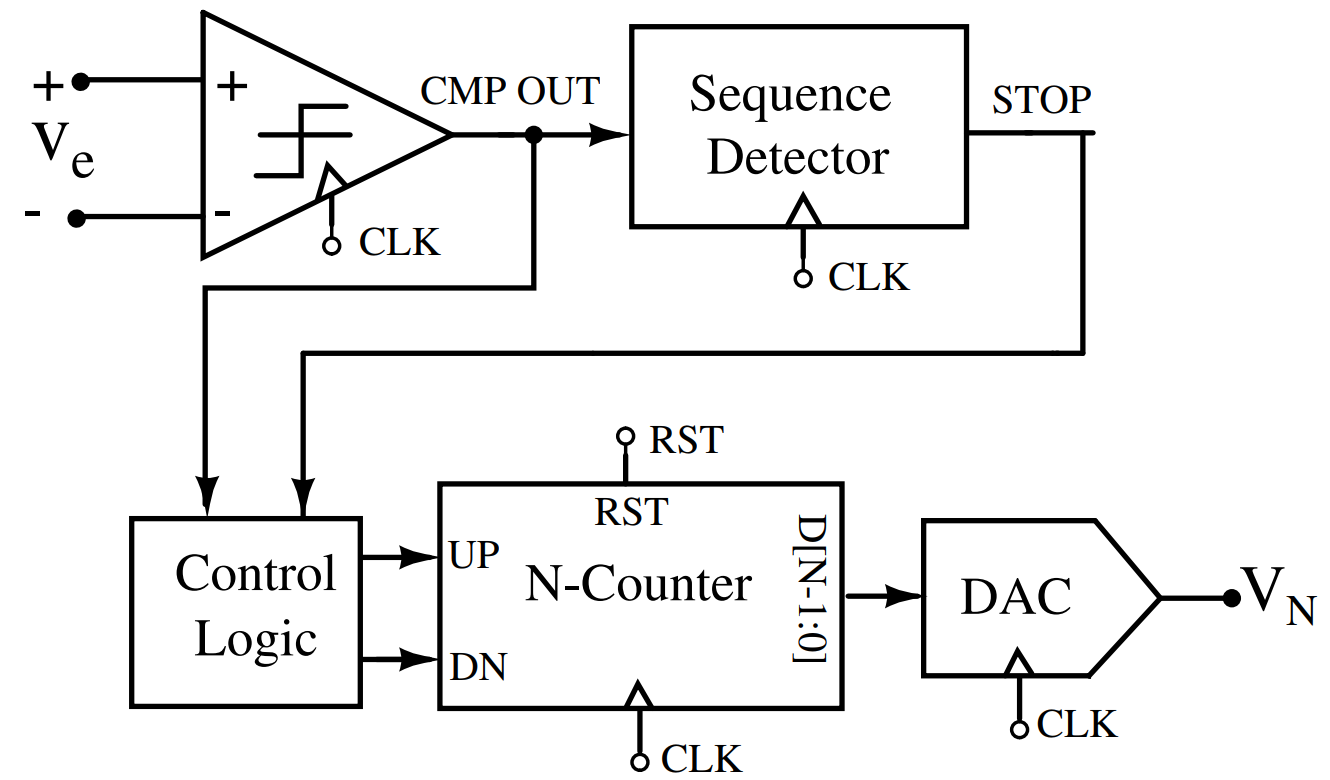}
   \caption{Internal architecture of $V_N$ detection block~\cite{marinaik2023}.}
   \label{2.4}
   \vspace{0.3cm}
\end{figure}

\begin{figure}[h!]
   \centering
   \includegraphics[scale=0.13]{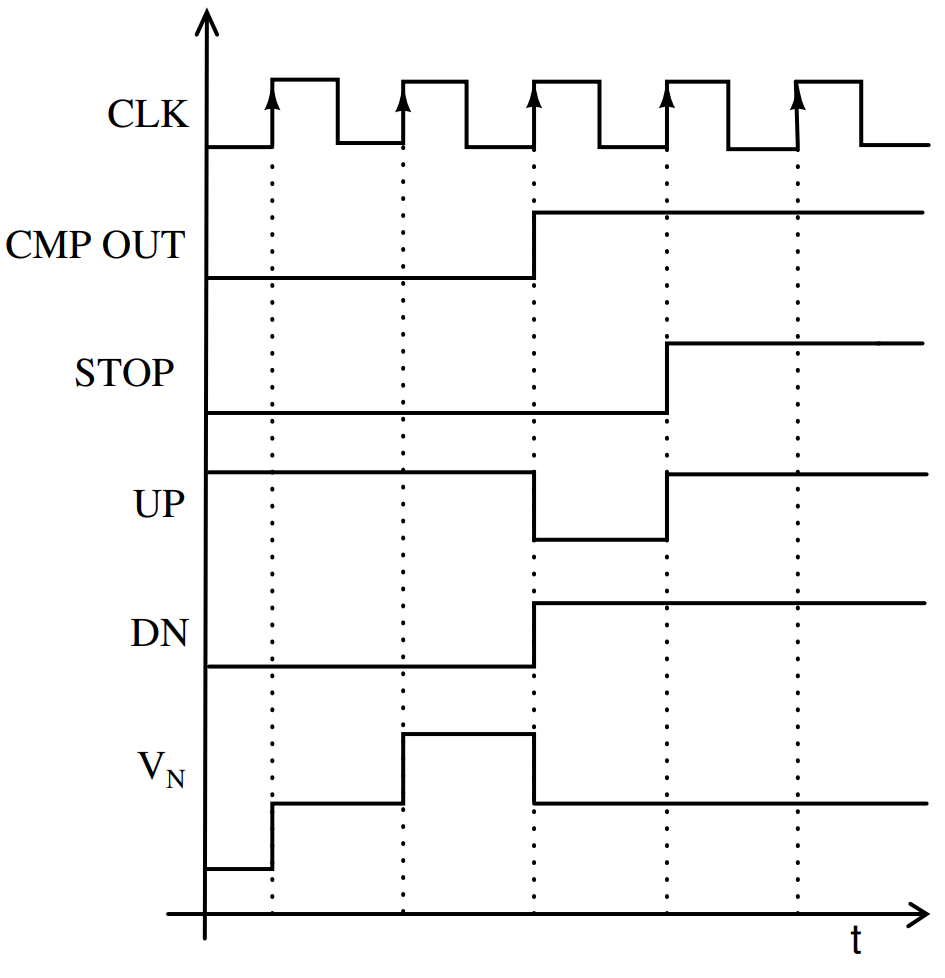}
   \caption{Timing diagram showing various signal transitions with DAC settling to $V_N$ \cite{marinaik2023}.}
   \label{2.5}
   \vspace{0.2cm}
\end{figure}

It is clear that the steady-state error $v_e$ becomes zero if the net conductance connected at the OTA load, i.e., $g_{ON}$, becomes zero. As shown in equation (1), this implies $g_o = g_N$, meaning the output conductance is cancelled. However, this condition requires very high gain error amplifiers ($E_{A}$), which are practically challenging \cite{marinaik2023}. In this technique, digital circuitry is used to obtain the value of $V_N$. The block-level diagram and the corresponding timing diagram are shown in Fig.~\ref {2.4} and Fig.~\ref {2.5}, respectively~\cite{marinaik2023}.

\begin{figure}[h!]
   \centering
   \includegraphics[scale=0.38]{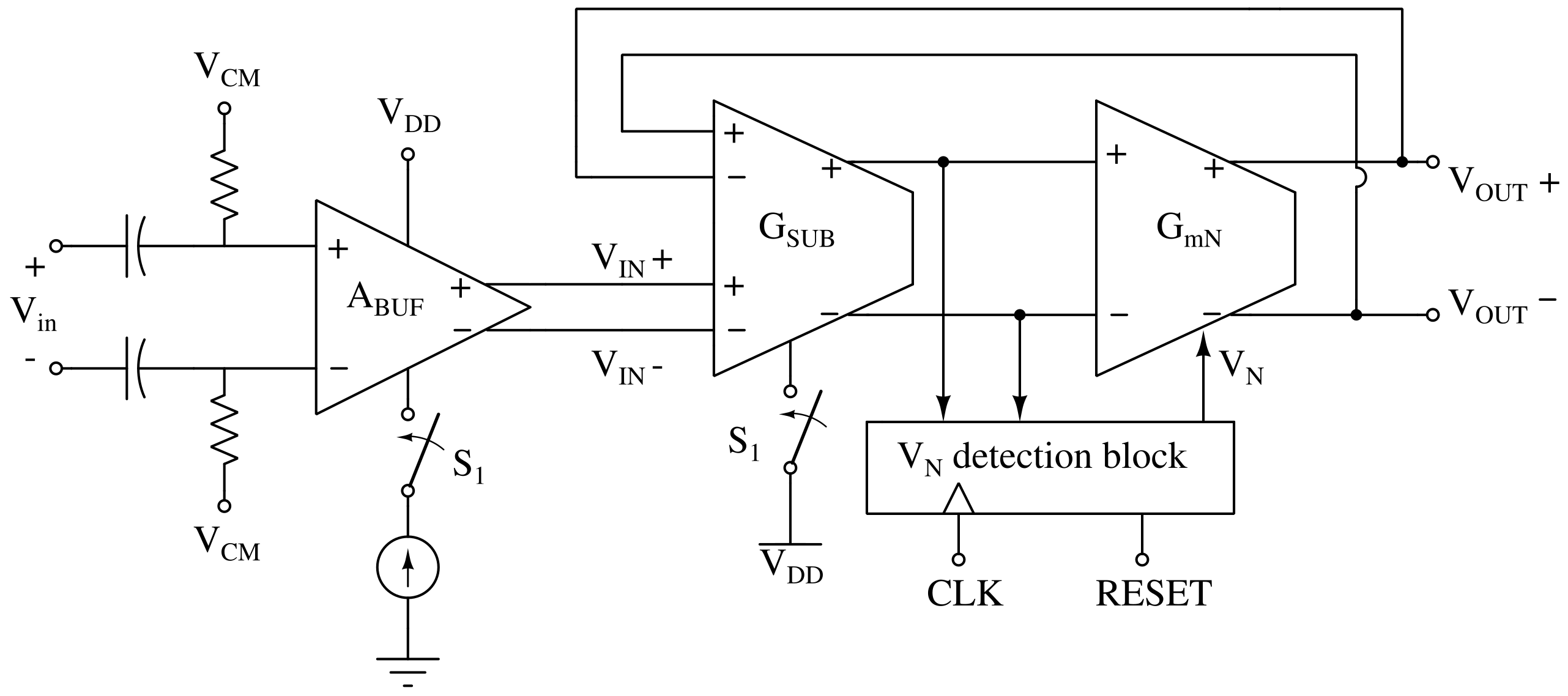}
   \caption{Block-level representation of high-gain OTA operating in different modes of operation: Calibration mode (when switch S\textsubscript{1} is OFF) and Normal mode (when switch S\textsubscript{1} is ON). \cite{marinaik2023}}
   \label{2.6}
   \vspace{0.2cm}
\end{figure}

Fig.~\ref {2.6} shows the overall block-level representation of the OTA implemented with this technique, where the OTA is configured in different operating modes.

The OTA gain calibration is performed using the feedback-based technique described above, in which the DAC control code is iteratively adjusted until the error voltage approaches zero, after which the calibrated OTA is used in normal operation.

The practical implementation challenges associated with this technique, including comparator design and its motivation for and implementation of offset calibration, DAC implementation, and the effects of subtractor gain and offset introduction, are discussed in Section~II, followed by the corresponding simulation results in Section~III, followed by the conclusions in Section~IV.


\section{Implementation Challenges}
\subsection{Motivation and Implementation of Comparator Calibration}
For an OTA with a DC gain of 20~dB, the steady-state error voltage $v_e$ is approximately 18~mV for a 200~mV incremental input. To achieve a DC gain of 50~dB, this error must be reduced to approximately 0.63~mV. In the $V_N$ detection block, the first stage is a clocked comparator; hence, the comparator must reliably detect input differences below 0.63~mV. However, the intrinsic offset of a clocked comparator is typically on the order of tens of millivolts making it unsuitable for the required calibration accuracy. Therefore, the comparator offset must be calibrated before OTA calibration.

A strong-arm latch-based Sense Amplifier (SA)\cite{razavi2015strongarm,razavi2020comparator}, with the improved slave latch reported in~\cite{nikolic2000}. is used as the comparator, with a clock frequency of 1 MHz and a power supply voltage Vdd of 1.2~Volts, and the design is optimized for minimum offset. \nkst{The working principle and design of the SA comparator are discussed in detail in \cite{razavi2015strongarm} and \cite{razavi2020comparator}, respectively. Instead of a NAND latch, a back-to-back inverter-based latch \cite{nikolic2000} is used because of its capability to produce both the outputs $Out$ and $\overline{OUT}$ almost at the same time, i.e., with very little relative delay. }

\begin{figure}[h!]
   \vspace{0.5cm}
   \centering
   \includegraphics[scale=0.45]{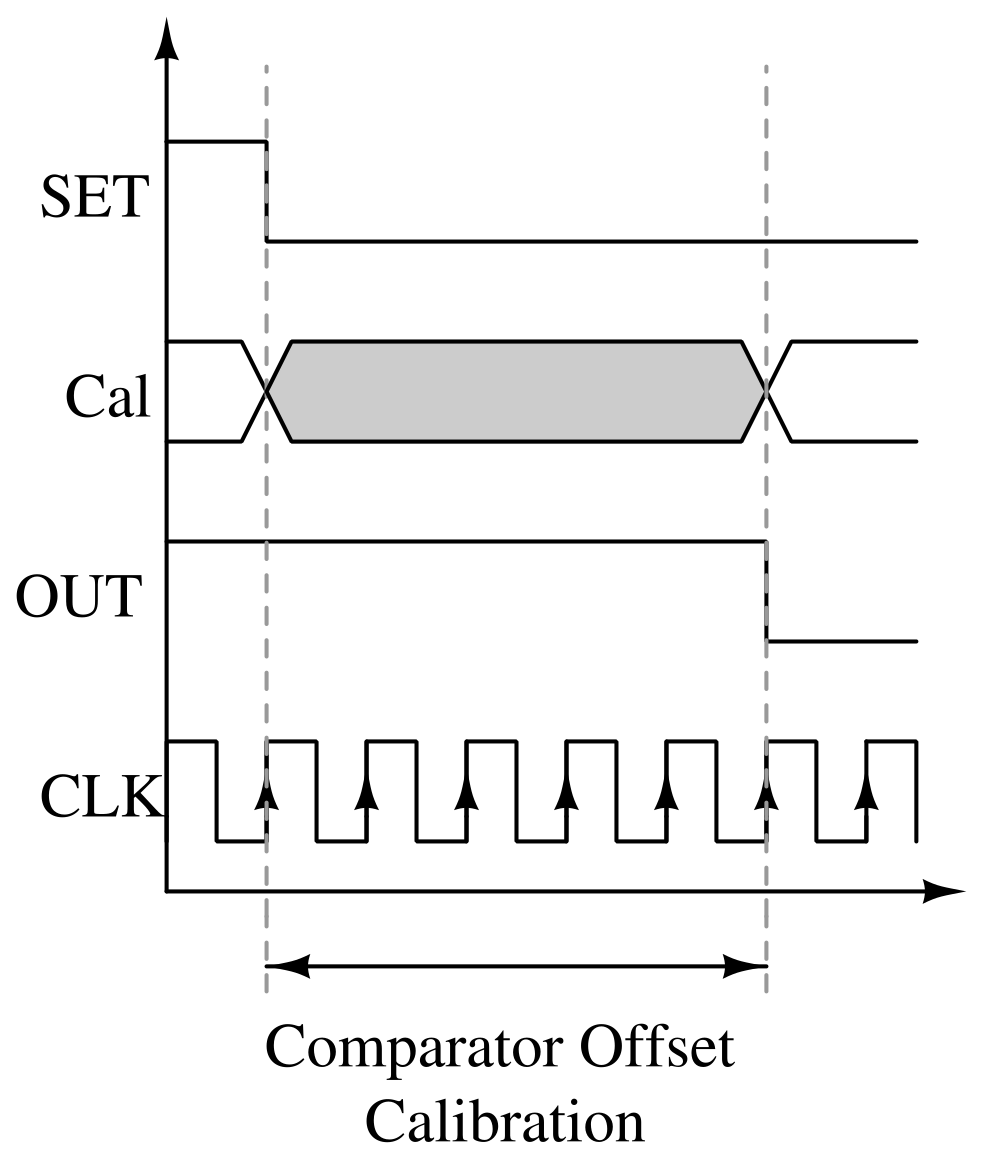}
   \caption{Timing diagram showing the SET signal to initialize the output to logic high and calibration for offset starts when the SET signal switches to low.}
   \label{3}
   \vspace{0.0cm}
\end{figure}

\begin{figure}[h!]
   \vspace{0.5cm}
   \centering
   \includegraphics[scale=0.38]{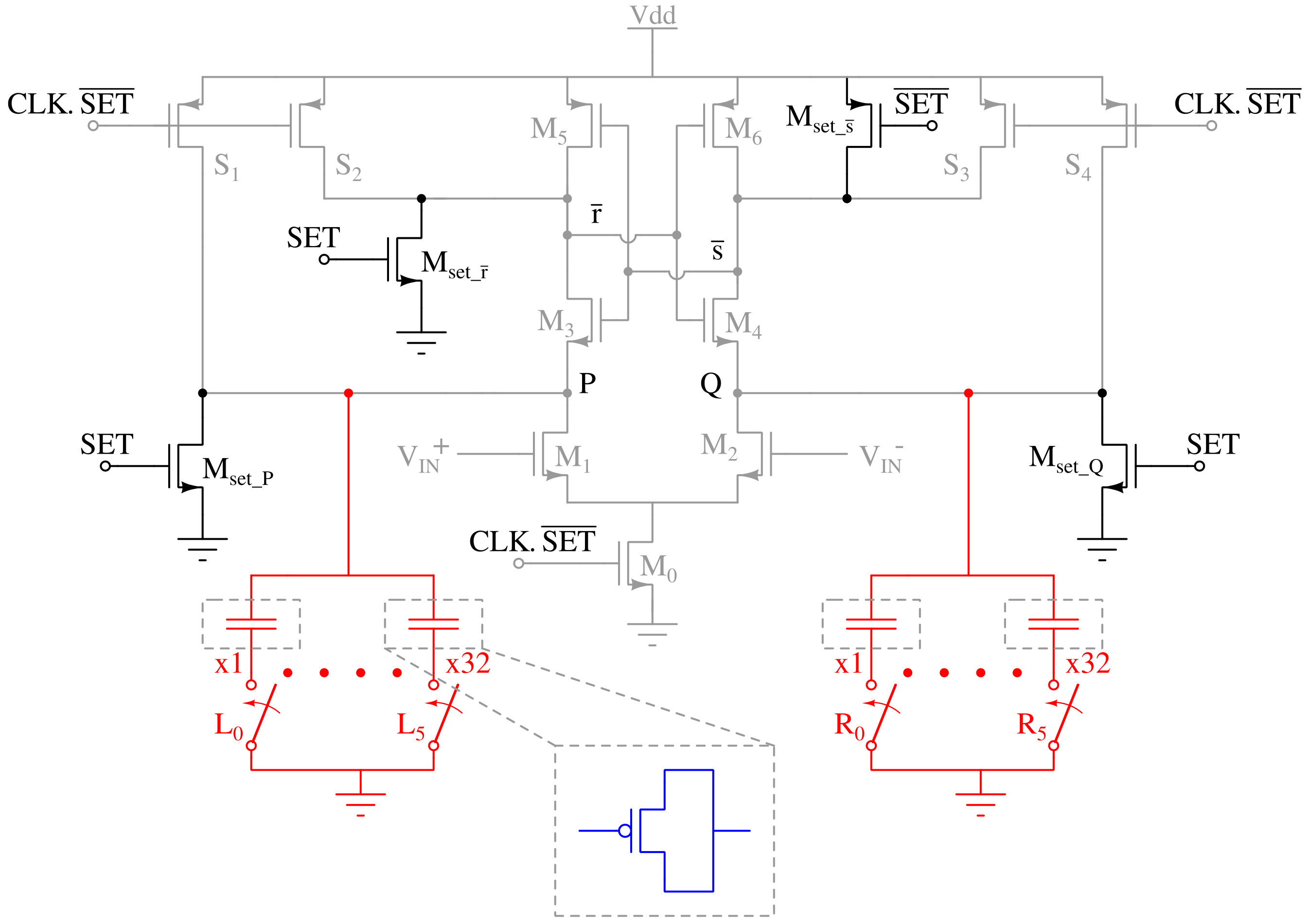}
   \caption{Proposed set scheme for sense-amplifier based comparator by setting the $\overline{s}$ node and resetting the $\overline{r}$, P, and Q nodes connected to its capacitor array (red- unit capacitance: 1.04~fF) realized by a MOSCAP (blue- unit MOSCAP dimensions: $2\mu$m/$60$nm).}
   \label{3.7}
   \vspace{0.0cm}
\end{figure}

A binary-weighted capacitive DAC (CDAC), shown in Fig.~\ref{3.7}, is implemented in this work to calibrate the comparator offset. The in-line calibration technique also requires the comparator to provide a SET functionality, which forces its output to a known logic-high state independent of the input signals. Conventional masking achieves this by forcing the internal nodes $\overline{r}$ and $\overline{s}$ to logic low and high, respectively. However, the asymmetric circuits used to drive these nodes introduce unequal capacitive loading, which can result in additional offset. To minimize this effect, the proposed sense amplifier (SA) employs a balanced push-pull SET mechanism. Dedicated transistors $M_{set\_\overline{s}}$, $M_{set\_\overline{r}}$, $M_{set\_P}$, and $M_{set\_Q}$, referred to as output-setting MOSFETs, force nodes $P$, $Q$, and $\overline{r}$ to ground and $\overline{s}$ to $V_{\mathrm{DD}}$, as shown in Fig.~\ref{3.7}. The widths of these transistors are appropriately sized to balance the loading of the internal nodes and minimize the resulting offset. When $\mathrm{SET}=1$, the comparator output is forced to logic high; when $\mathrm{SET}$ transitions to logic low, the forced state is released, and the comparator resumes normal operation. The resulting SET operation and subsequent comparator operation during calibration are shown in Fig.~\ref{3}.

\begin{figure}[h!]
   \vspace{0.0cm}
   \centering
   \includegraphics[scale=0.17]{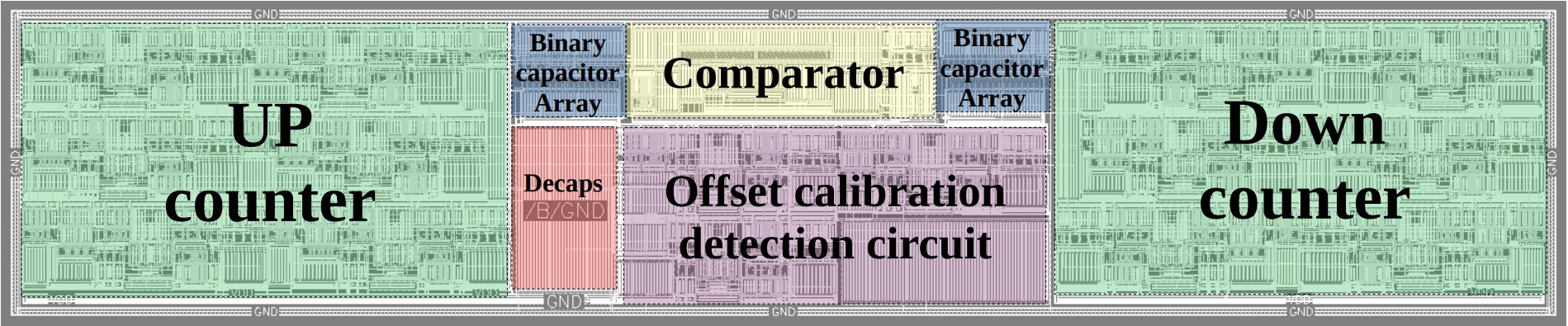}
   \caption{Complete comparator with calibration circuitry layout highlighting the up and down counters in green, comparator offset calibration detection circuitry in purple, decaps in red, comparator in yellow, and the binary capacitor arrays in blue.}
   \label{3.9}
   \vspace{0.0cm}
\end{figure}

In the implemented comparator's offset calibration method, the offset generated by the binary CDAC is initially steered entirely to one side, and a linear search is done to find the correct calibration code. During initialization, the entire binary-weighted CDAC is connected to the drain of the input NMOS of the comparator while the CDAC at the opposite input transistor's drain is left disconnected. With the first clock edge, the comparator output is forced into a known deterministic state, assuming the CDAC-induced offset is larger than the intrinsic comparator offset. In the next cycle, the smallest unit capacitor of the CDAC is connected to the previously disconnected input, thereby reducing the net differential offset applied to the comparator. Subsequent clock edges alternately remove one unit of capacitance from the initially driven side and add one unit to the opposite side, effectively performing a linear search toward zero offset. After each adjustment, the comparator output is monitored; when it switches state, the applied CDAC offset is taken to be the compensation value for the intrinsic offset. The calibration loop is then immediately terminated, and the corresponding CDAC control code is latched. The incremental and decremental switching of CDAC elements required for this procedure is managed by digital counters, which hold the calibration code. Fig.~\ref{3.9} shows the layout of the comparator, including the offset calibration technique, highlighting SA, latch, CDAC, offset feedback control unit, and counters in different colours.


\begin{figure}[h!]
   \vspace{0cm}
   \centering
   \includegraphics[scale=0.06]{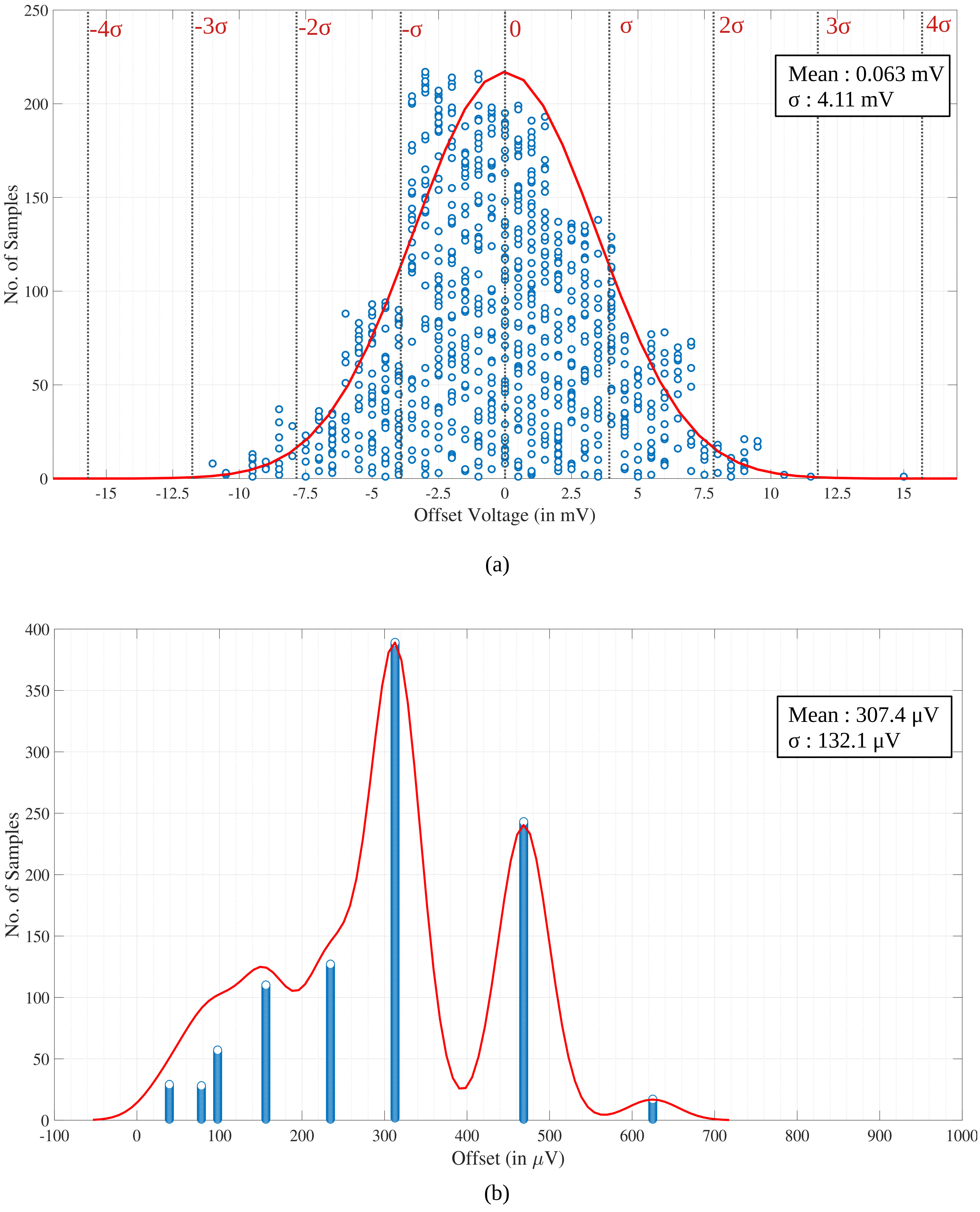}
   \caption{Post-Layout Monte-Carlo simulation results before (a) and after (b) offset calibration of the comparator, indicating that all the samples have an offset less than 0.9 mV.}
   \label{3.10}
   \vspace{0.3cm}
\end{figure}

The comparator is evaluated using 1000 Monte Carlo mismatch samples. Before calibration, the measured standard deviation is $\sigma = 4.11$~mV, corresponding to a $4\sigma$ worst-case offset of 16.44~mV, as shown in Fig.~\ref{3.10}(a). Since the comparator offset must be below 1.19~mV to achieve the required minimum OTA gain of 60~dB, offset calibration is necessary. The post-calibration results are shown in Fig.~\ref{3.10}(b). Unlike the pre-calibration distribution, the calibrated samples are no longer randomly distributed around the nominal offset and therefore do not follow a Gaussian distribution. Consequently, the mean and $4\sigma$ values are not directly representative of the post-calibration distribution. Instead, the post-calibration offset is evaluated directly from the simulated samples. The resulting mean offset is 307~$\mu$V, with a standard deviation of 132~$\mu$V and a worst-case offset of 625~$\mu$V, all of which are below the required 1.19~mV limit. With the comparator offset sufficiently reduced, the gain-calibration loop can subsequently be activated to calibrate the OTA gain. 

\subsection{DAC}
\begin{figure}[h!]
    \vspace{0cm}
	\centering
	\includegraphics[scale=0.4]{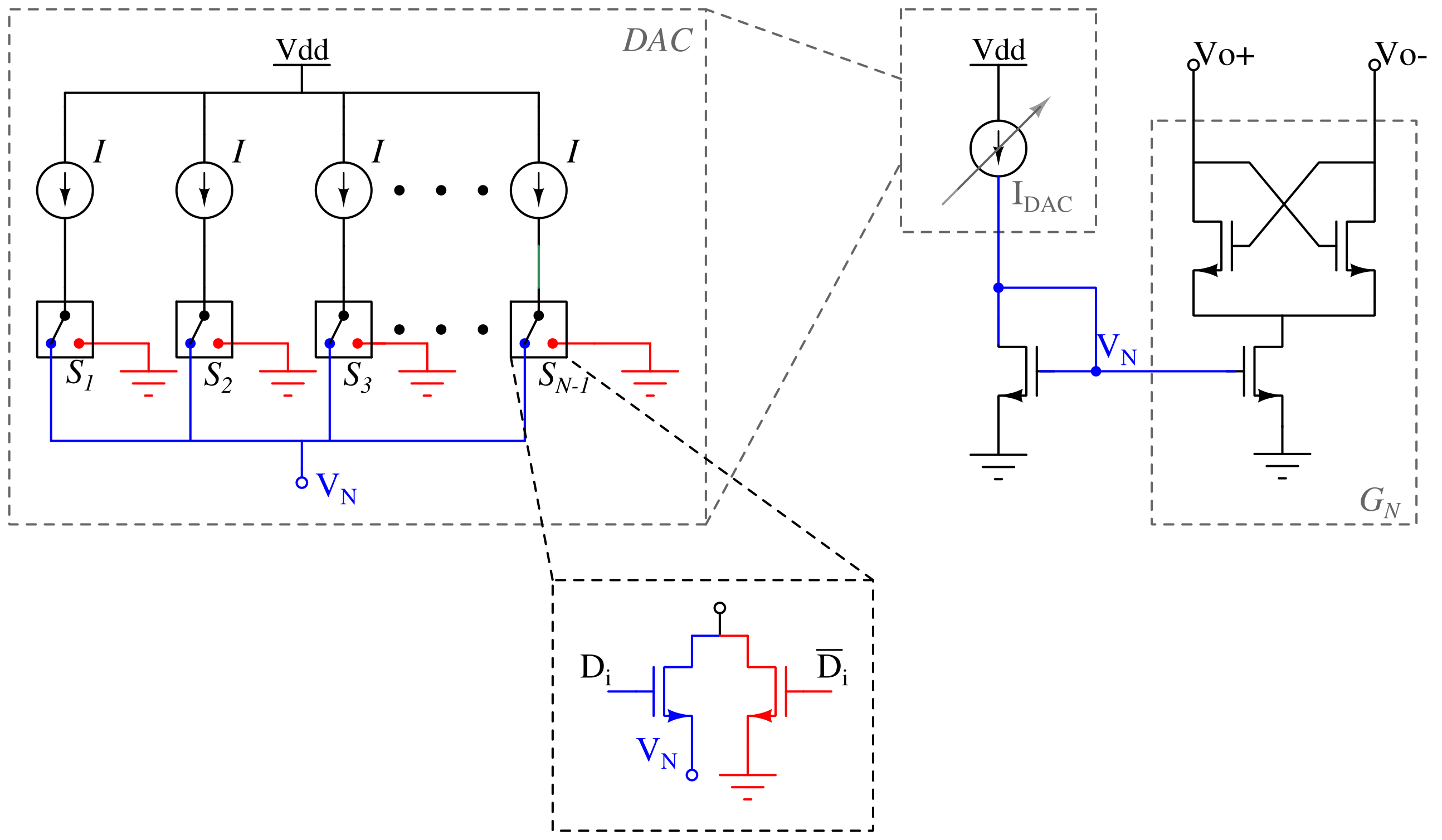}
	\caption{Unary I-DAC for biasing of positive-feedback transconductor $G_N$. The binary to thermometer-converted counter outputs $D_i$ are the input for DAC switches, enabling the current(s) $I$ to pass to either ground(Red) or contribute to $V_N$ (Blue).}
	\label{4.2}
    \vspace{0cm}
\end{figure}

A 4-bit thermometer-coded current-steering DAC, shown in Fig.~\ref{4.2}, converts the binary counter output into the analog control voltage $V_{N}$ for the tail MOSFET of the cross-coupled pair (X-CP). The counter output is first converted to thermometer code to ensure monotonicity, which then controls an array of $2^{n}-1$ identical p-type MOS current sources. Each branch uses two  NMOS switches to steer its current either to the diode-connected MOSFET or to ground. The selected branch currents are summed at the diode-connected MOSFET to generate $V_{N}$, which also serves as the reference device for the X-CP tail-current mirror.

\subsection{Subtractor gain and offset generation}

Let us revisit the system-level architecture shown in Figs.~\ref{2.3} and~\ref{2.6}. The subtractor generates the error voltage $\Delta v_e$ by subtracting the output voltage $\Delta v_o$ from the applied offset $\Delta v_i$, where the forward-path relation is given by

\begin{equation}
\Delta v_o = A_{ON}\cdot A_{SUB}\cdot v_e,
\end{equation}

where $A_{ON}\cdot A_{SUB}$ is the overall forward-path gain. 


Thus, the subtractor gain must be accounted for when specifying the required OTA gain. For example, a subtractor with a gain of 10~dB contributes directly to the 60~dB overall gain, leaving 50~dB for the OTA. Likewise, with a 10~dB subtractor, if we were to target an OTA gain of 60~dB OTA, it results in a 70~dB forward-path gain, requiring $v_e<316~\mu$V rather than 1.19~mV for the same calibration accuracy. This directly tightens the required comparator offset and, consequently, the comparator calibration range and resolution. In this work, a unity gain subtractor is used. 

\nkst{Therefore, a low-gain subtractor is preferred; in this work, the subtractor gain is 0.02~dB, allowing an OTA gain of 59.98~dB for an overall forward-path gain of approximately 60~dB.}

\begin{figure}[h!]
    \vspace{0cm}
	\centering
	\includegraphics[scale=0.55]{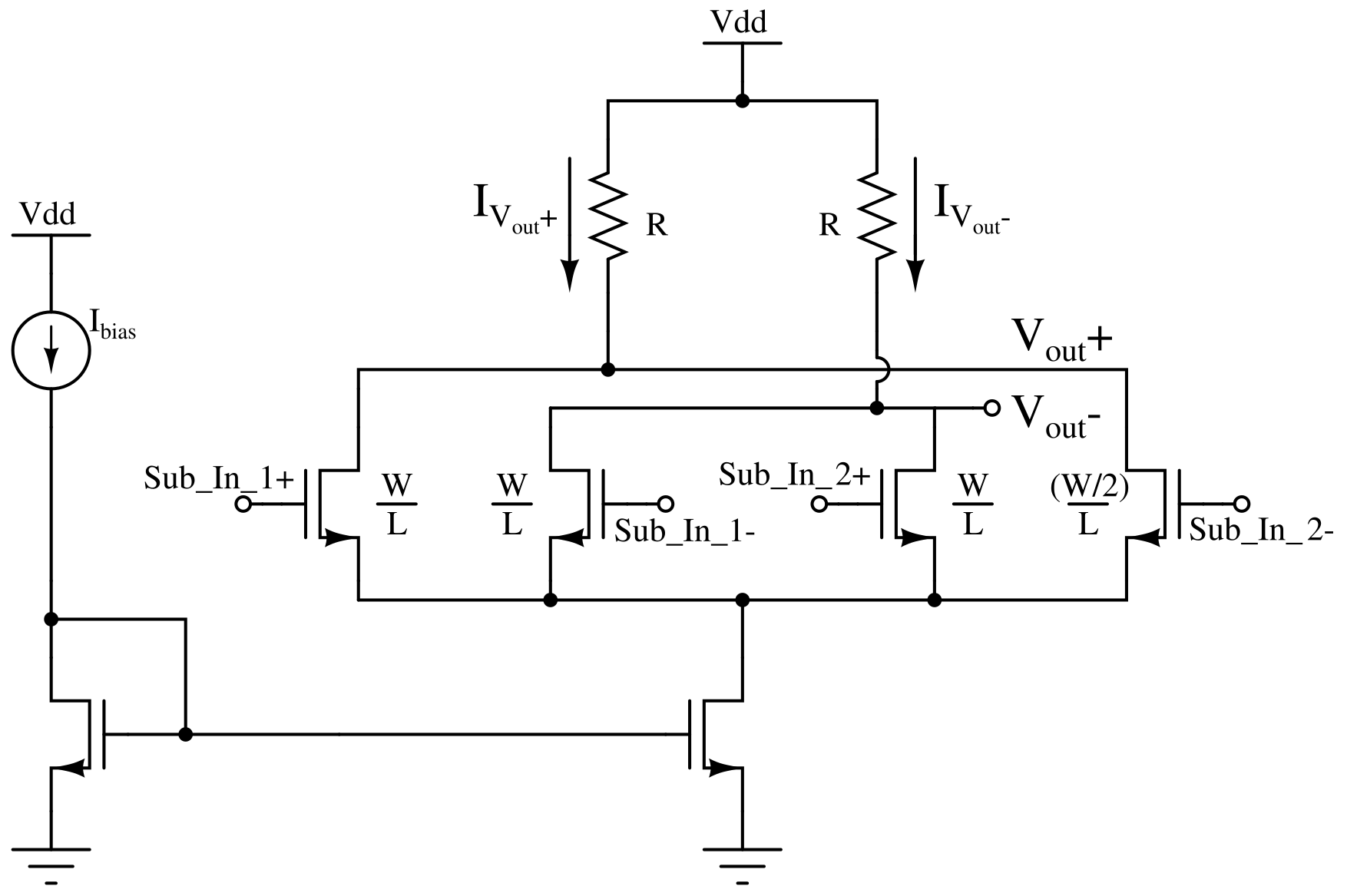}
	\caption{Subtractor implementation with intentionally sized NMOS input pairs for systematic offset generation during OTA gain calibration.}
	\label{4.12}
    \vspace{0cm}
\end{figure}

A systematic offset, \(\Delta v_i\), is required for OTA gain calibration and is generated in the subtractor by intentionally introducing an imbalance between the currents \(I_{V_{\mathrm{out}+}}\) and \(I_{V_{\mathrm{out}-}}\) flowing through the resistors \(R\), as shown in Fig.~\ref{4.12}. In the fully differential implementation, this imbalance is achieved by introducing a mismatch in the geometry of the subtractor's input transistors. The resulting effective incremental input voltage, \(\Delta v_i\), is approximately \(180\,\mathrm{mV}\). Assuming a maximum allowable residual error of \(625\,\mu\mathrm{V}\), the calibration relationship estimates an effective OTA gain of approximately \(51\,\mathrm{dB}\).

\begin{table*}[h!]
\centering
\caption{Comparison of the OTA used in this work with previously reported architectures.}
\label{tab:OTA-comparision}

\resizebox{\textwidth}{!}{%
\begin{tabular}{|c|c|c|c|c|c|c|}
\hline
\textbf{Parameters} 
& \shortstack{\textbf{OTA without}\\\textbf{impedance}\\\textbf{cancellation}} 
& \textbf{This work} 
& \textbf{\cite{mondal2015}} 
& \textbf{\cite{Gain2}} 
& \textbf{\cite{nauta1992}} 
& \textbf{\cite{Gain1}} \\ \hline

Technology node (nm) 
& 65 
& 65 
& 130 
& 350 
& 300 
& 1000 \\ \hline

Supply Voltage (V) 
& 1.2 
& 1.2 
& 1.2 
& 3 
& 10 
& 2 \\ \hline

DC Gain (dB) 
& 18 
& 60 
& 48 
& 83 
& 46 
& 60 \\ \hline

Power (mW) 
& 0.16 
& 0.45 
& 9 
& 2.4 
& 670 
& 102 \\ \hline

UGB (GHz) 
& 10 
& 10 
& 20 
& 0.13 ($C_L=2$ pF) 
& 10 
& 0.84 \\ \hline

Immune to PVT 
& No 
& Yes 
& Yes 
& No 
& No 
& No \\ \hline

Immune to mismatch 
& No 
& Yes 
& No 
& No 
& No 
& No \\ \hline

Manual / Automatic calibration 
& Not Applicable 
& Automatic 
& Automatic 
& Manual (Off-chip) 
& Manual (Off-chip) 
& Manual (Off-chip) \\ \hline

\end{tabular}
}
\end{table*}

\section{Post-Layout Simulation Results}

\begin{figure}[h!]
    \vspace{0cm}
	\centering
	\includegraphics[scale=0.25]{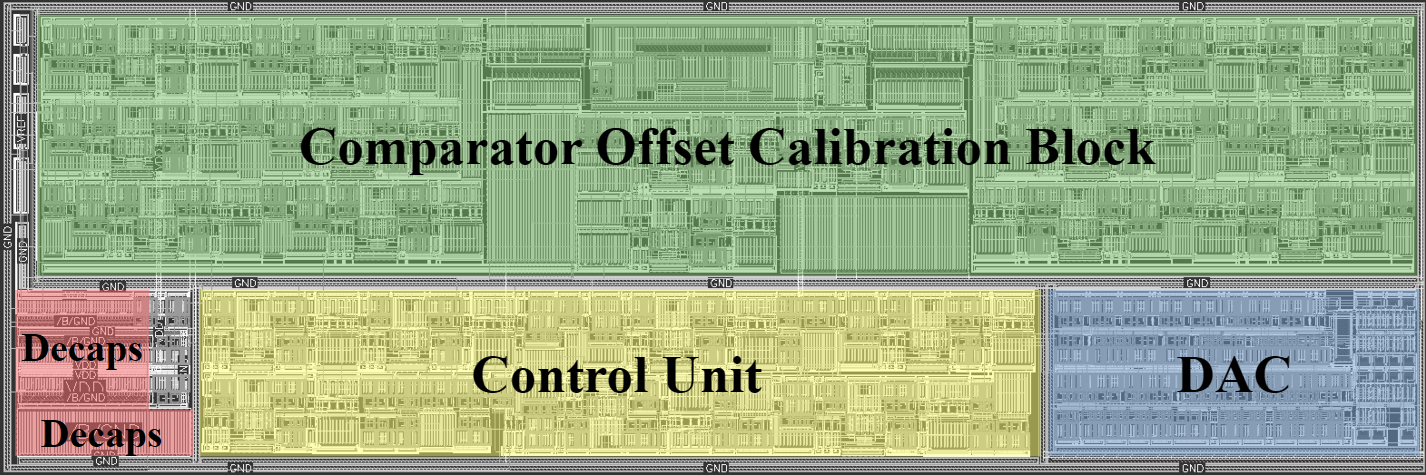}
	\caption{Layout of the comparator calibration circuit integrated with the gain calibration circuit, highlighting the Control unit as well as DAC.}
	\label{4.7}
    \vspace{0cm}
\end{figure}

\nkst{
Using the comparator before its offset correction isn't meaningful; consequently, making the technique as well as the OTA of no use. Hence, calibrating the comparator offset is the first step. Also, it can be said that the overall system goes through two calibrations: one is the calibration for the comparator offset, and the second one is the OTA gain calibration.}
\nk{As discussed in the previous section, the calibration routine includes offset calibration followed by gain calibration.} During offset calibration, both the inputs of the comparator will be tied to some common voltage $V_{CM}$. As soon as calibration is done, the comparator gets connected to the digital loop, and its input will be connected to the error voltage $v_{e}$. Hence, a few clock cycles will be required first for comparator calibration, and then in the next few cycles, the main OTA will be calibrated. Fig.~\ref{4.7} depicts the layout of the comparator offset calibration circuit integrated with the gain calibration circuit. The complete layout is 142 $\mu$m $\times$ 47 $\mu$m.

\begin{figure}[h!]
    \vspace{0.1cm}
	\centering
	\includegraphics[scale=0.09]{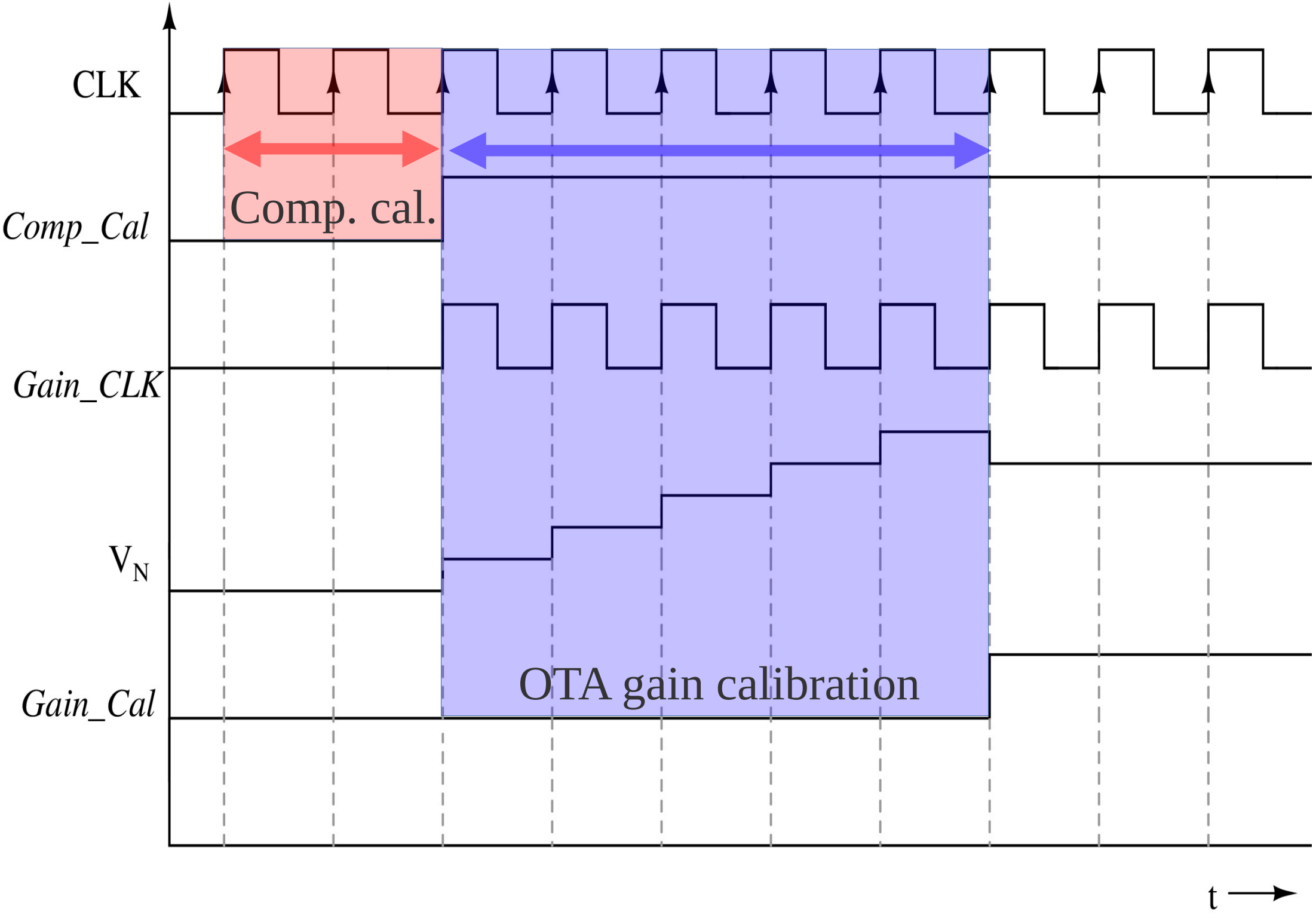}
	\caption{Timing diagram showing the calibration signal of the comparator (Red) as well as the OTA (Blue).}
	\label{4.9}
    \vspace{0cm}
\end{figure}

The timing diagram is depicted in Figure~\ref{4.9}. For illustration only, comparator calibration is shown as being done in 2 clock cycles. Once the comparator is calibrated, the signal Comp\_Cal switches from 0 to 1, and the gated clock, Gain\_CLK, is supplied to the main digital circuitry of the $V_{N}$ - detection block; which then calibrates the main OTA for its gain. Similarly, once the gain is calibrated, the signal Gain\_cal switches to 1. During gain calibration, the OTA needs to be in calibration mode \cite{marinaik2023}, meaning the incremental input voltage $\Delta$v$_{i}$ needs to be active during this time. Once the gain is calibrated, $\Delta$v$_{i}$ gets disconnected, and the subtractor will be configured as a buffer.


\section{Conclusion}

This work presented a practical transistor-level implementation of a high-gain OTA calibration technique, with emphasis on the circuit-level challenges. A comparator offset calibration scheme based on a 6-bit capacitive DAC was implemented in 65-nm CMOS to enable reliable detection of the sub-millivolt error required for high OTA gain. Post-layout Monte Carlo simulations demonstrated a standard deviation of 132~$\mu$V and a maximum observed offset of 625~$\mu$V after calibration, compared with a pre-calibration standard deviation of 4.11~mV and a $4\sigma$ offset of 16.44~mV.\nkst{ The implemented DAC provides a calibration range of 24~mV, exceeding the required 19~mV range. In addition, the gain-calibration architecture incorporates a subtractor with a gain of 1 to minimize its contribution to the overall forward-path gain and the resulting error-detection requirement.} The complete calibration circuitry, including the comparator calibration, digital control logic, DAC, and OTA gain-calibration circuitry, was \nkst{physically }implemented and evaluated using post-layout simulations. The results demonstrate the feasibility of integrating the calibration technique at the circuit level while addressing practical constraints such as comparator offset, DAC implementation, signal-path loading, and layout-related parasitics. The resulting calibrated high-gain OTA provides the foundation for its integration into larger analog systems requiring accurate and robust transconductance and gain characteristics.

\section{Acknowledgment}

The authors would like to thank Mayur S. Marinaik, for valuable discussions and constructive feedback that contributed to this work. 




\begin{thebibliography}{00}


\bibitem{1} B. Razavi, ``Design of Analog CMOS Integrated Circuits'', 2nd ed. McGraw-Hill Education, 2005, indian student version.
\bibitem{2} Y. Taur and T. H. Ning, Fundamentals of modern VLSI devices. Cambridge University Press, 2021.
\bibitem{marinaik2023} M. S. Marinaik and N. Kadayinti, “Mismatch tolerant negative conductance load tuning for high gain OTAs,” in IEEE Asia Pacific Conference On Circuits and Systems (APCCAS), 2023, pp. 363–366.
\bibitem{nauta1992} B. Nauta, “A CMOS transconductance-C filter technique for very high frequencies,” IEEE Journal of Solid-State Circuits, vol. 27, no. 2, pp. 142–153, Feb. 1992.
\bibitem{Gain1} S. Xiao and C. Salama, “Voltage gain enhancement by conductance cancellation in GaAs MESFET op-amps,” in 1993 IEEE International Symposium on Circuits and Systems (ISCAS), 1993, pp. 1073–1076 vol.2.
\bibitem{Gain2} J. Yan and R. Geiger, “A negative conductance voltage gain enhancement technique for low voltage high speed CMOS op amp design,” in Proceedings of the 43rd IEEE Midwest Symposium on Circuits and Systems (Cat.No.CH37144), vol. 1, 2000, pp. 502–505 vol.1.
\bibitem{mondal2015} I. Mondal and N. Krishnapura, “Gain enhanced high frequency ota with on-chip tuned negative conductance load,” in IEEE International Symposium on Circuits and Systems (ISCAS), 2015, pp. 2085–2088.
\bibitem{razavi2015strongarm} B. Razavi, “The strongarm latch [a circuit for all seasons],” IEEE Solid-State Circuits Magazine, vol. 7, no. 2, pp. 12–17, 2015.
\bibitem{razavi2020comparator} B. Razavi, “Design of comparator [the analog mind],” IEEE Solid-State Circuits Magazine, vol. 12, no. 4, pp. 8–14, 2020.
\bibitem{nikolic2000} B. Nikolic, V. G. Oklobdzija, V. Stojanovic, W. Jia, J. K.-S. Chiu, and M. M.-T. Leung, “Improved sense-amplifier-based flip-flop: Design and measurements,” IEEE Journal of Solid-State Circuits, vol. 35, no. 6, pp. 876–884, Jun. 2000.




\end{thebibliography}
\end{document}